# Pressure dependence of the dielectric loss
# in semi-conducting polypyrrole aged at room temperature

**I. Sakellis, A.N. Papathanassiou and J. Grammatikakis**

*University of Athens, Physics Department, Section of Solid State Physics, Panepistimiopolis, 15784 Zografos, Athens, Greece*

## ABSTRACT

The effect of physical aging of semi-conducting polypyrrole at ambient temperature for two years duration on the dielectric loss at various pressures is investigated. Changes of the dielectric loss spectra and the modification of the values of the activation volume for relaxation are interpreted through the division of chain clusters into smaller components and the reduction of the size of the conductive grains.

The electric charge transport in conducting polypyrrole proceeds by polaron hopping through two distinct conduction modes: intra-chain and inter-chain hopping. This picture can be generalized, by considering highly conducting clusters of chains, forming conducting grains embedded in an insulating environment [1]. Conductive grains consist of polymer chains rich in polarons [2]. To a crude approximation, metallic conduction occurs within the grain and hopping from one grain to another proceeds by phonon assisted tunneling through the insulating separation. While dc electrical conductivity measurements probe long range polaron transport whereas the higher potential barriers dominate and inter-chain transfer dominate, ac conductivity experiments can probe polaron motion of different scales depending on the frequency of the applied electric field [3, 4, 5, 6, 7, 8]. Pressure can enhance the ability of ac schemes to activate electric charge flow along pathways of different lengths. In a recent publication, the ac conductivity spectrum of conducting polypyrrole was splitted in a pressure sensitive and a pressure insensitive frequency regions attributed to inter and intra-chain hopping respectively [9, 10]. By manipulating the ac conductivity data into complex permittivity ones, it was found that a broad dielectric loss mechanism detected at ambient pressure, was gradually decomposed into two constituents on increasing the pressure: one relaxation mechanism depends weakly on pressure (intra-chain relaxation) and another one which is pressure dependent (inter-chain relaxation). The first (pressure-insensitive) mechanism was attributed to intra-chain (or intra-grain) relaxation, as chains are less compressible than the whole polymer (chains plus void space). The second (pressure sensitive) one yields negative activation volume of polaron relaxation. This was interpreted as an inwards relaxation of the specimen whilst polaron passes through the void space separating two polymer chains [3, 4]. In this approach, to a first approximation, the density of electric charge carriers is regarded as pressure-independent, in the pressure range our experiments are performed.

In this work, we report room temperature complex impedance measurements at various pressures (from ambient pressure to 300 MPa) in semi-conducting polypyrrole aged at room temperature for two years. We stress that, in the present work, the term aging refers to physical aging and no chemical aging occurs. Details about the sample preparation were reported earlier [11]. The present results are compared with those obtained two years before the current results [9, 10]. A Novocontrol (Germany) high- pressure dielectric spectroscopy system (0.3 GPa is the maximum hydrostatic pressure), which used silicone oil as a pressure transmitting fluid. Metallic electrodes were pasted on to the parallel surfaces of the specimen and a very thin insulating layer of epoxy covered the specimen to prevent contamination from the pressure transmitting fluid [12, 13, 14, 15].

The imaginary part of the relative permittivity ε″ (i.e., the permittivity reduced to the free space permittivity) is connected to the real part of the complex conductivity σ′ as:

$$\varepsilon''(\omega) = \frac{\sigma'(\omega)}{\varepsilon_0 \omega} = \frac{\sigma_0}{\varepsilon_0 \omega} + \varepsilon_d'' \qquad (1)$$

where ω denotes the angular frequency (ω=2πf; f is the frequency of the applied harmonic electric field), $\varepsilon_0$ is the permittivity of free space, $\sigma_0$ is the frequency-independent conductivity (usually labeled as the dc conductivity) and $\varepsilon_d''$ is the imaginary part of the relative permittivity, after subtraction of the dc component:

$$\varepsilon_d''(\omega) = \frac{\sigma'(\omega) - \sigma_0}{\varepsilon_0 \omega} \qquad (2)$$

The activation volume $\upsilon^{act}$ for relaxation is defined as $\upsilon^{act} \equiv \left( \partial g^{act} / \partial P \right)_T$, where $g^{act}$ denotes the Gibbs free energy for activation. From the isothermal pressure evolution of the dielectric loss peak, we can determine the value of the activation volume $\upsilon^{act}$ for polaron relaxation [10].

$$\upsilon^{act} \cong -kT \left( \frac{\partial \ln f_{max}}{\partial P} \right)_T \qquad (3)$$

where $f_{max}$ is the frequency where $\varepsilon_d''(f)$ has a maximum. The activation volume is the volume fluctuation induced in a hopping procedure. Inspired by the definition given in Ref. [16], we can state that, for inter-chain hopping, it is the specimen's volume change when the hopping polaron is instantly located at the inter-chain void space (excited state) compared with that when the polaron is cited on a chain (ground state) [9, 10]: passage of a polaron through the void space yields reduction of the distortion and the sign of the corresponding activation volume is expected to be negative [9, 10].

The imaginary part of the relative permittivity $\varepsilon_d''(f)$ (i.e., after subtraction of the dc component $\sigma_0$ from the measured $\sigma(\omega)$ values, according to Eq. (2)) of aged polypyrrole is presented in Figure 1, together with results from the same specimen recorded two years earlier. We label I and II the low and high frequency peak, respectively. The effect of aging on these mechanisms is the following:

(i) Relaxation I is shifted to lower frequency and gets more intense
(ii) Relaxation II is slightly shifted toward higher frequency and its intensity gets suppressed

The variation of $\ln f_{max}(P)$ for relaxation II is shown in the inset of Figure 2. A straight line fitted to the experimental data points, yields, by employing Eq. (3), the activation volume for relaxation II: $v_{II}^{act} = (-3.7 \pm 0.6) \,\text{Å}^3$. We note that relaxation I is practically pressure-insensitive. Analysis of the experimental results gives $v_{I}^{act} = (-0.1 \pm 1.0) \,\text{Å}^3$; the experimental error prohibits an accurate knowledge of the activation volume, which is very small and very close to zero.

Aging of semi-conducting polypyrrole induces various changes to the inhomogeneous structure of the polymer. The size of the conductive grains decreases, the disorder increases and the density of counter-ions is suppressed. As a result, electric charge transport is hindered after () aging. Inter-chain (or inter-grain) transport is prohibited because polymer chains break into smaller pieces and create additional trap sites, which act as traps to the intra-chain motion. Additionally, intra-chain conductivity is suppressed due to the reduction of the density of counter-ions that act as 'links' to hopping in conducting polymer systems. Thus, mechanism I requires longer relaxation times and subsequently to the shift of the maximum toward lower frequency. The latter is indeed observed experimentally (see (i)). The reason that mechanism I gets intensified with aging (see (i)) is that – as mentioned above – the division of chains to smaller parts increases the population of dead-ends that act as traps (in combination to the reduction of the density of counter-ions linking hopping sites) and, subsequently intra-chain localization is pronounced.

aging makes inter-chain transport difficult, because of the reduction of size of conductive grains and the decrease of the counter-ions that serve as 'bridges' between inter-chains hopping [17]. Fewer carriers are capable of performing inter-chain hopping and, subsequently, the intensity of relaxation II (which is proportional to the density of polarons that participate in this process) gets decreased with aging (see (ii)). The location of relaxation II is not significantly modified with aging (see (ii)), indicating that inter-chain chain is not seriously affected by aging, which seems quite reasonable (aging modifies the heterogeneous structure, but the typical inter-chain distance is not affected).

The *absolute* value of the activation volume for relaxation II is reduced with aging (e.g., the present results should compare with the corresponding value of the activation volume for non-aged polypirrole: $(-6.9 \pm 0.3) \, \text{Å}^3$) [8]). This seems quite reasonable since relaxation II proceeds by inter-chain (or, inter-grain) hopping along distances that get modified by the breaking of chains and the reduction of the size of the conductive grains. $\upsilon_{II}^{act}$ is not a localized volume fluctuation, but is an average over the whole volume of the specimen: aging modifies the stability of polarons and is related with the loss of counter-ions. Transverse polarons and bipolarons extend over two or more of chains and over cross-linked chain aggregates. Aging reduces polarons to one-dimensional ones, which are more stable, but have limited contribution to charge transport. At the same time, the loss of counter-ions from the bulk prohibits the existence of transverse polarons [18].

At this point, we stress that the negative sign of the activation volume for relaxation is thermodynamically consistent. Consider the elastic thermodynamic model, known as the cBΩ model [19 20, 21, 22], which asserts that Gibbs free energy for defect formation, migration or activation is related dominantly to the elastic work required for each one of the aforementioned processes and is proportional to the isothermal bulk modulus, i.e., $g^{act} = cB\Omega$, where $g^{act}$ denotes the Gibbs free energy for activation ($g^{act} = h^{act} - Ts^{act}$, where $h^{act}$ and $s^{act}$ are the activation enthalpy and the activation entropy, respectively), c is practically constant and Ω is a volume related with the mean atomic volume. The validity of this model has been checked at ambient pressure in various classes of solids extending from silver halides [22] to rare gas solids [23] as well as in ionic crystals under gradually increasing uniaxial stress [24] in which electric signals are emitted before fracture (in a similar fashion as the electric signals detected before earthquakes [25, 26, 27, 28]). Under By differentiating with respect to pressure, we get $\upsilon^{act} = B^{-1}[(\partial B/\partial P)_T - 1]g^{act}$ ., where $\gamma_i$ is the Grüneisen mode responsible for relaxation ($\gamma \equiv -(\partial \ln \nu / \partial \ln V)_T$, where ν and V denote a phonon frequency and volume, respectively). For a simple mono-atomic solid, and within the Debye approximation, this equation can take the form $\upsilon^{act} = 2\gamma_1 g^{act}/B$ [29]. For ionic crystals, the later can reduce to: .. [29] $\upsilon^{act} = 2(\gamma - 1/3)g^{act}/B$. The negative sign of the activation volume can probably result if the specific Grüneisen parameter involved with the relaxation process is less than 1/3.

As a conclusion, long-term physical aging of semi-conducting polypyrrole yields changes of the dielectric loss spectra and to the value of the activation volume of inter-chain (or inter-grain) polaron hopping. The results are interpreted through the reduction of the size of the conductive grains, the division of polymer chains into smaller ones and the decrease of the density of counter-ions.

**Figure captions**

Figure 1. The (relative) imaginary part of the permittivity $\varepsilon_d''$ (obtained after subtraction of the dc component) at 300MPa vs frequency. The () aging duration between these measurements was 2 years.

Figure 2. Typical curves of the imaginary part of the relative permittivity $\varepsilon_d''$ vs frequency of aged polypyrrole at room temperature and various pressures. From top to bottom: ambient pressure, 100 MPa, 200 MPa and 300 MPa. The pressure evolution of $f_{max}$ is depicted in the inset. Straight lines are best fits to the data points.

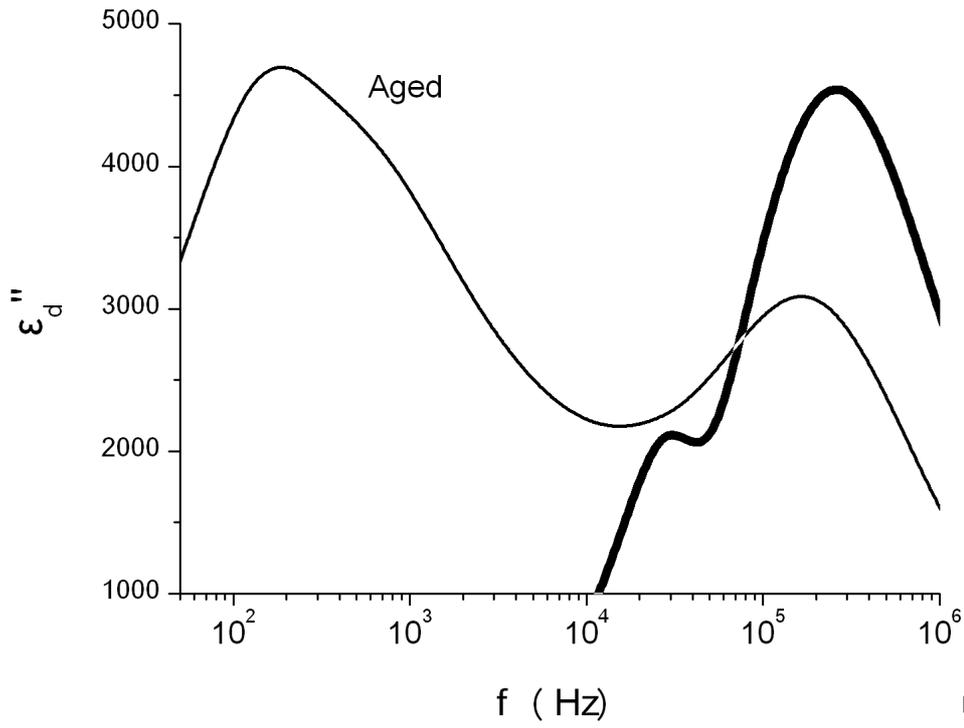

FIGURE 1

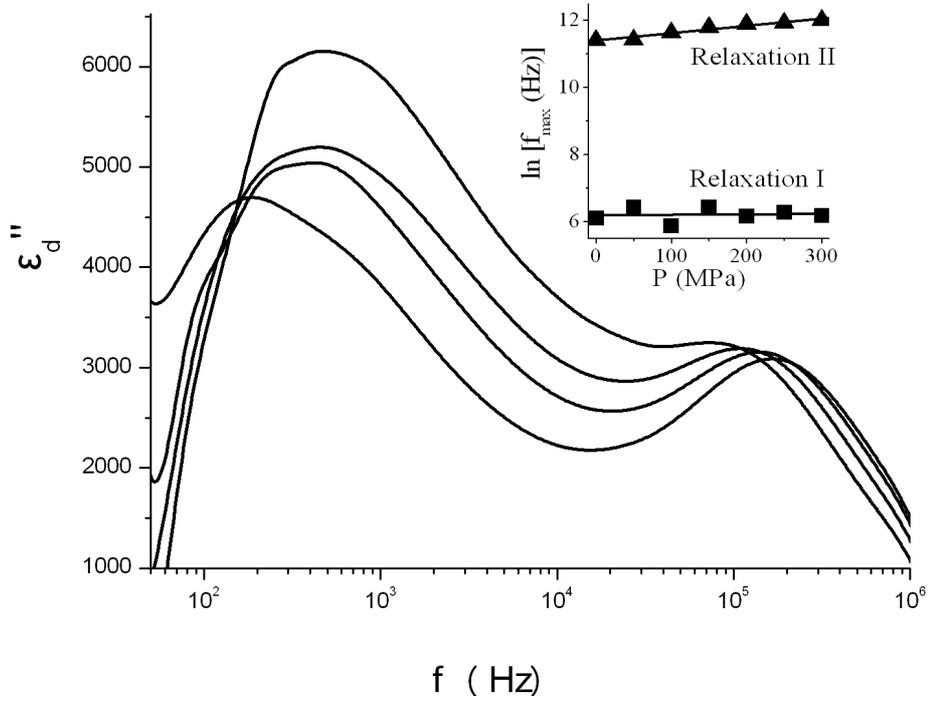

FIGURE 2